\documentclass[twocolumn]{aastex63}

\shorttitle{Phase Modeling of TRAPPIST-1}
\shortauthors{Stephen R. Kane et al.}

\usepackage{amsmath}
\usepackage{multirow}

\begin{document}

\title{Phase Modeling of the TRAPPIST-1 Planetary Atmospheres}

\author[0000-0002-7084-0529]{Stephen R. Kane}
\affiliation{Department of Earth and Planetary Sciences, University of
  California, Riverside, CA 92521, USA}
\email{skane@ucr.edu}

\author[/0000-0002-8444-3436]{Tiffany Jansen}
\affiliation{Department of Astronomy, Columbia University, New York,
  NY 10027, USA}

\author[0000-0002-5967-9631]{Thomas Fauchez}
\affiliation{NASA Goddard Space Flight Center, Greenbelt, MD 20771,
  USA}

\author[0000-0001-9619-5356]{Franck Selsis}
\affiliation{Laboratoire d'astrophysique de Bordeaux, Univ. Bordeaux,
  CNRS, B18N, all\'ee Geoffroy Saint-Hilaire, 33615 Pessac, France}
  
\author[0000-0002-0438-1803]{Alma Y. Ceja}
\affiliation{Department of Earth and Planetary Sciences, University of
  California, Riverside, CA 92521, USA}

%%%%%%%%%%%%%%%%%%%%%%%%%%%%%%%%%%%%%%%%%%%%%%%%%%%%%%%%%%%%%%%%%%%%

\begin{abstract}

Transiting compact multi-planet systems provide many unique
opportunities to characterize the planets, including studies of size
distributions, mean densities, orbital dynamics, and atmospheric
compositions. The relatively short orbital periods in these systems
ensure that events requiring specific orbital locations of the planets
(such as primary transit and secondary eclipse points) occur with high
frequency. The orbital motion and associated phase variations of the
planets provide a means to constrain the atmospheric compositions
through measurement of their albedos. Here we describe the expected
phase variations of the TRAPPIST-1 system and times of superior
conjunction when the summation of phase effects produce maximum
amplitudes. We also describe the infrared flux emitted by the
TRAPPIST-1 planets and the influence on the overall phase
amplitudes. We further present the results from using the global
circulation model {\sc ROCKE-3D} to model the atmospheres of
TRAPPIST-1e and TRAPPIST-1f assuming modern Earth and Archean
atmospheric compositions. These simulations are used to calculate
predicted phase curves for both reflected light and thermal emission
components. We discuss the detectability of these signatures and the
future prospects for similar studies of phase variations for
relatively faint M stars.

\end{abstract}


%%%%%%%%%%%%%%%%%%%%%%%%%%%%%%%%%%%%%%%%%%%%%%%%%%%%%%%%%%%%%%%%%%%%

\section{Introduction}
\label{intro}

The phase variations of an exoplanet are caused by the observed
reflected light and thermal emission components of an exoplanet as it
orbits the host star. Reflected phase signatures can reveal the albedo
and scattering properties of planetary atmospheres and thus provide a
unique insight into their compositions, while thermal signatures can
reveal the efficiency of heat transport from the dayside to the
nightside. Though small in amplitude compared with the flux of the
star, the cyclic nature of the phase variations allow them to be
detected given a long enough temporal baseline of observations. In
particular, precision photometry from the {\it Kepler} mission has
demonstrated that phase signatures of orbiting exoplanets are indeed
present and detectable in the data
\citep[e.g.,][]{esteves2013,esteves2015,quintana2013,angerhausen2015b}. Such
phase investigations are continuing in the era of the Transiting
Exoplanet Survey Satellite ({\it TESS}), with phase signatures
detected for numerous known exoplanets
\citep[e.g.,][]{shporer2019,jansen2020,wong2020b}.

The recent discovery of the TRAPPIST-1 planetary system presents an
interesting opportunity to study how atmospheric reflectivity and heat
transport can affect the photometric observations. The system was
initially detected by \citet{gillon2016} and found to harbor three
terrestrial transiting planets. Continued monitoring of the system
with ground and space-based observatories revealed that the system has
four additional terrestrial transiting planets \citep{gillon2017a},
three of which are within the Habitable Zone (HZ) of the host star
\citep{bolmont2017a}, with planet e having the most potentially
favorable conditions for surface liquid water
\citep{wolf2017b}. Analysis of the Transit Timing Variations (TTVs) by
\citet{grimm2018} produced improved mass and density estimates,
constraining the interior models and fractions of volatiles. A lack of
absorption features from Hubble Space Telescope (HST) observations
during transit excludes cloud-free hydrogen-dominated atmospheres for
most of the planets, leaving open the potential for temperate surface
conditions \citep{dewit2018,ducrot2018,zhang2018}. Even so, the
activity of the host star and high XUV irradiation of the planets may
have had a profound effect on their atmospheres
\citep{becker2020a,peacock2019a,fleming2020}, possibly leading to
substantial loss of their atmospheric mass
\citep{roettenbacher2017,wheatley2017,hori2020}. Further observations
by the {\it K2} mission enabled the confirmation of the outer planet
and verification of its orbital period \citep{luger2017b}. A
compilation of transit data by \citet{agol2020b} further refined the
planetary masses to a precision of 3--5\%. The orbital architecture of
this compact planetary system ensures that there are relatively
frequent ``syzygy'' events, such as planet-planet occultations
\citep{luger2017c}, and occasions when multiple planets simultaneously
contribute to the maximum reflected light at superior
conjunction. Such events will allow tests of atmospheric models to be
conducted based on the amplitude of the phase signatures and the
passbands at which they are observed.

In this paper we model the phase variations of the TRAPPIST-1 system
and the connection to models of the planetary atmospheres. In
Section~\ref{theory} we summarize the theoretical methodology to
derive the photometric phase variations. In Section~\ref{comb} we
calculate the phase variations of the TRAPPIST-1 planetary system and
predict maximum phase amplitudes for the individual planets and
combined phase events for various geometric albedo assumptions. The
phase variations resulting from global circulation models (GCM) of
TRAPPIST-1e and TRAPPIST-1f are described in Section~\ref{gcm},
including short-wave reflected light and long-wave thermal emission
components. Section~\ref{discussion} contains a discussion of
detectability prospects using current and future facilities, and the
importance of distinguishing between different atmospheric evolution
scenarios. Finally, we provide a summary of our work and concluding
remarks in Section~\ref{conclusions}.

%%%%%%%%%%%%%%%%%%%%%%%%%%%%%%%%%%%%%%%%%%%%%%%%%%%%%%%%%%%%%%%%%%%%

\section{Photometric Phase Variations}
\label{theory}

Photometry of exoplanet host stars with sufficient photometric
precision and observational cadence can reveal the phase variations due
to the planets \citep{shporer2017a} and can distinguish between
planetary and stellar companions \citep{kane2012b}. Here we describe
the variations at optical wavelengths due to the reflected and
scattered light components of the photons received from the
planet. Due to the relatively low mass of the planets in the
TRAPPIST-1 system, the Doppler beaming and ellipsoidal variation
components have a negligible effect and are discussed in
Section~\ref{discussion}.

Here we adopt the formalism of \citet{kane2010b,kane2011a}. The flux
ratio of a planet with radius $R_p$ to the host star at wavelength
$\lambda$ and phase angle $\alpha$ is given by
\begin{equation}
  \epsilon(\alpha,\lambda) \equiv
  \frac{f_p(\alpha,\lambda)}{f_\star(\lambda)}
  = A_g(\lambda) g(\alpha,\lambda) \frac{R_p^2}{r^2}
  \label{fluxratio}
\end{equation}
where $A_g(\lambda)$ is the geometric albedo and $g(\alpha,\lambda)$
is the phase function. The star--planet separation, $r$, is given by
\begin{equation}
  r = \frac{a (1 - e^2)}{1 + e \cos f}
  \label{separation}
\end{equation}
where $a$ is the semi-major axis, $e$ is the orbital eccentricity, and
$f$ is the true anomaly. The phase angle, defined to be zero when the
planet is at superior conjunction, is given by
\begin{equation}
  \cos \alpha = - \sin (\omega + f)
  \label{phaseangle}
\end{equation}
The phase function $g(\alpha,\lambda)$ is often approximated by a
Lambert sphere, which assumes the atmosphere isotropically scatters
over $2 \pi$ steradians. Here we adopt the empirically derived
``Hilton function'' \citep{hilton1992}, based upon observations of
Jupiter and Venus and represented as a visual magnitude correction of
the form
\begin{equation}
  \Delta m (\alpha) = 0.09 (\alpha/100\degr) + 2.39
  (\alpha/100\degr)^2 -0.65 (\alpha/100\degr)^3
  \label{magcorr}
\end{equation}
resulting in a phase function of the form
\begin{equation}
  g(\alpha) = 10^{-0.4 \Delta m (\alpha)}
  \label{phase}
\end{equation}
One of the main measurables from the detection of phase variations is
the geometric albedo $A_g(\lambda)$, which in turn depends upon the
surface conditions of the planet. Since many of the planets in the
TRAPPIST-1 system are terrestrial and complete atmospheric desiccation
may have occurred, there are various possible surface scenarios. For
example, \citet{kane2011f} describe the three basic scenarios of
rock, molten, and atmosphere in the context of 55~Cancri~e, where rock
and molten correspond to geometric albedos of 0.1 and 0.6
respectively. Given the equilibrium temperatures of the TRAPPIST-1
planets \citep{gillon2017a}, they are unlikely to have molten
surfaces, but we adopt a "reflective" surface/atmosphere to represent
the high geometric albedo of 0.6. Geometric albedos for planets with
atmospheres vary enormously, depending on composition, cloud decks,
haze layers, etc \citep{jansen2018,madden2018,mansfield2019}. For the
atmosphere scenario, we adopt an Earth geometric albedo of 0.434,
particularly as it lies in between the bare rock and reflective
surface scenarios described above.

%%%%%%%%%%%%%%%%%%%%%%%%%%%%%%%%%%%%%%%%%%%%%%%%%%%%%%%%%%%%%%%%%%%%

\section{Combined Phase Amplitude}
\label{comb}

As noted in Section~\ref{intro}, the TRAPPIST-1 system is a
particularly interesting science case, partly due to the compact
nature of the system resulting in a high frequency of full orbital
phases for each of the planets. Such a system poses a modeling
challenge since the phase variations for all planets must be accounted
for \citep{kane2013b}, but can also be an advantage if the planets
regularly line up near superior conjunction where their phase
amplitudes combine for a stronger effect \citep{gelino2014}. This is
especially true for the TRAPPIST-1 system since the planets are close
to orbital resonance \citep{gillon2017a}, ensuring regular occurrence
of such superposition of phase effects.

The properties of the TRAPPIST-1 planets that are relevant to phase
variations are shown in Table~\ref{tab:flux}
\citep{gillon2017a,luger2017b}. All seven planets are within 0.06~AU
of the host star, with orbital periods all less than 20~days. The
sizes of the planets indicate that all of them are in the terrestrial
regime and, as mentioned in Section~\ref{intro}, three of the planets
lie within the conservative HZ and an additional planet lies within
the optimistic HZ (see \citet{kane2016c} for definitions of
conservative and optimistic HZ boundaries). A study of the effect of
revised stellar distances by \citet{kane2018a} found that the
TRAPPIST-1 planet semi-major axes and radii are relatively unaffected
by the {\it Gaia} DR2 release \citep{brown2018}.

\begin{deluxetable*}{ccccccccc}
  \tablecolumns{9}
  \tablewidth{0pc}
  \tablecaption{\label{tab:flux} TRAPPIST-1 planetary orbital
    parameters and flux ratios}
  \tablehead{
    \colhead{Planet} &
    \colhead{$P$} &
    \colhead{$a$} &
    \colhead{$R_p$} &
    \colhead{$i$} &
    \multicolumn{4}{c}{Flux Ratio (ppm)} \\
    \colhead{} &
    \colhead{(days)} &
    \colhead{(AU)} &
    \colhead{($R_\oplus$)} &
    \colhead{(deg)} &
    \colhead{Maximum} &
    \colhead{Rocky} &
    \colhead{Reflective} &
    \colhead{Atmosphere}
  }
  \startdata
  b & 1.5109 & 0.011 & 1.086 & 89.65 & 17.317 & 1.732 & 10.390 & 7.516 \\
  c & 2.4218 & 0.015 & 1.056 & 89.67 &  8.729 & 0.873 &  5.237 & 3.788 \\
  d & 4.0496 & 0.021 & 0.772 & 89.75 &  2.351 & 0.235 &  1.410 & 1.020 \\
  e & 6.0996 & 0.028 & 0.918 & 89.86 &  1.925 & 0.193 &  1.155 & 0.835 \\
  f & 9.2067 & 0.037 & 1.045 & 89.68 &  1.441 & 0.144 &  0.864 & 0.625 \\
  g & 12.353 & 0.045 & 1.127 & 89.71 &  1.132 & 0.113 &  0.679 & 0.491 \\
  h & 18.764 & 0.060 & 0.715 & 89.80 &  0.261 & 0.026 &  0.157 & 0.113 
  \enddata
  \tablecomments{Planetary properties extracted from
    \citet{gillon2017a} and \citet{luger2017b}.}
\end{deluxetable*}

\begin{figure*}
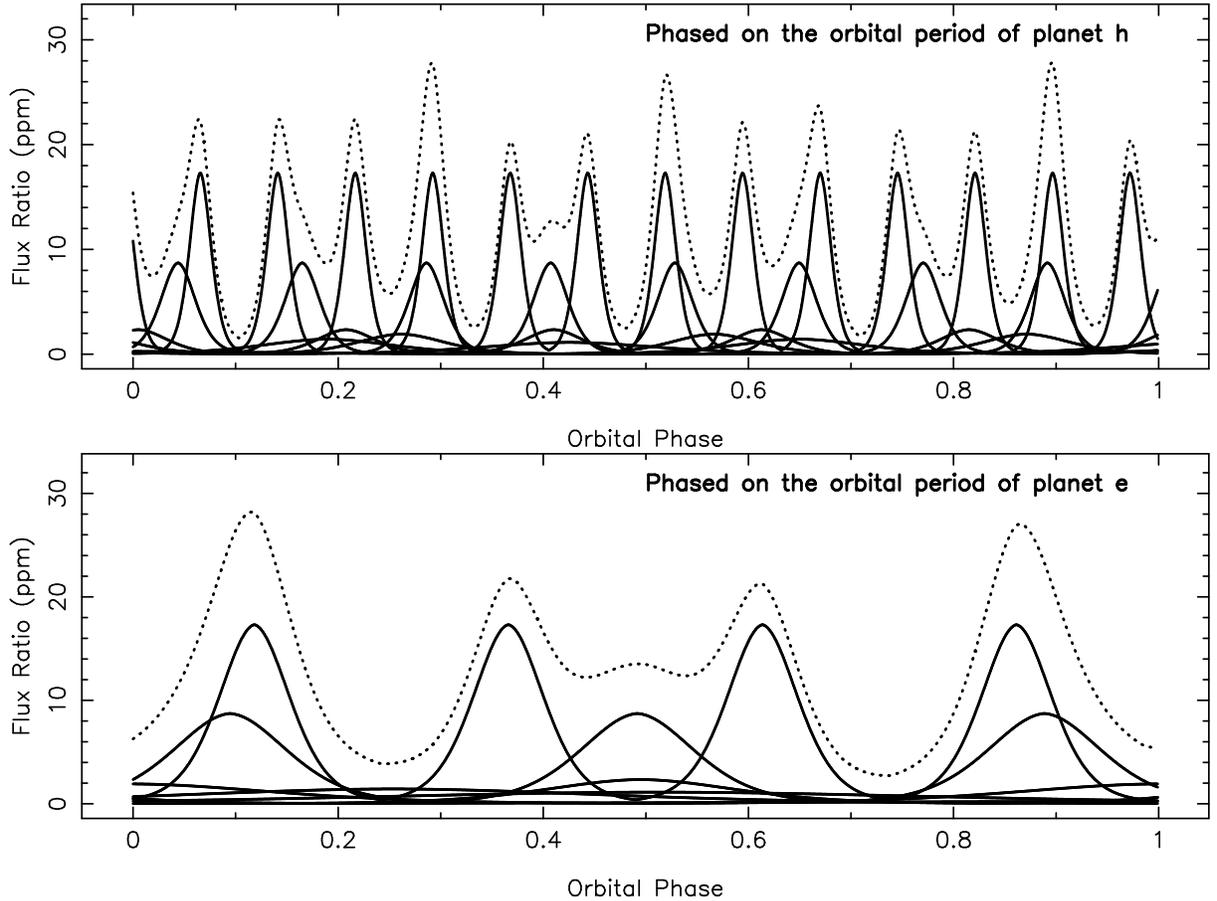

  \begin{center}
    \includegraphics[angle=270,width=16.0cm]{f01a.ps} \\
    \includegraphics[angle=270,width=16.0cm]{f01b.ps}
  \end{center}
  \caption{The combined flux variations of the TRAPPIST-1 system due
    to the reflected light from the planetary surfaces as a function
    of orbital phase. Individual planetary phase variations are shown
    are solid lines and the combined signature for all planets is
    shown as a dotted line. These calculations assume planetary
    albedos of unity so that the amplitudes may be scaled linearly to
    lower albedos. The top panel shows the phase variations for one
    complete orbit of the outermost planet, and the bottom panel is
    phased on the orbital period of planet e.}
  \label{fig:combined}
\end{figure*}

We use the methodology described in Section~\ref{theory} to construct
a predicted phase amplitude model as a function of time for the
system. As noted in Section~\ref{theory}, one of the primary
components in the model is the geometric albedo for the planets. The
phase amplitude for the three scenarios of rocky, reflective, and
atmosphere (see Section~\ref{theory}) are shown in
Table~\ref{tab:flux}, along with a "maximum" amplitude calculated for
an albedo of unity. These calculations include only the reflected
light component integrated over a broad (bolometric) passband. The
combined phase variation model using the unity albedos is represented
in Figure~\ref{fig:combined}, where the solid lines show the flux
ratio for the individual planets and the dotted line shows the
combined effect from all planets. The top panel is depicts the phase
variations for one complete orbit of the outer planet (h), and the
bottom panel shows these same variations on the scale of an orbit of a
HZ planet (e). As described by \citet{kane2013b}, an accurate orbital
ephemeris may be used to predict times when the combined effect of all
planets will produce a relatively high phase amplitude.

\begin{figure}
  \begin{center}
    \includegraphics[width=8.5cm]{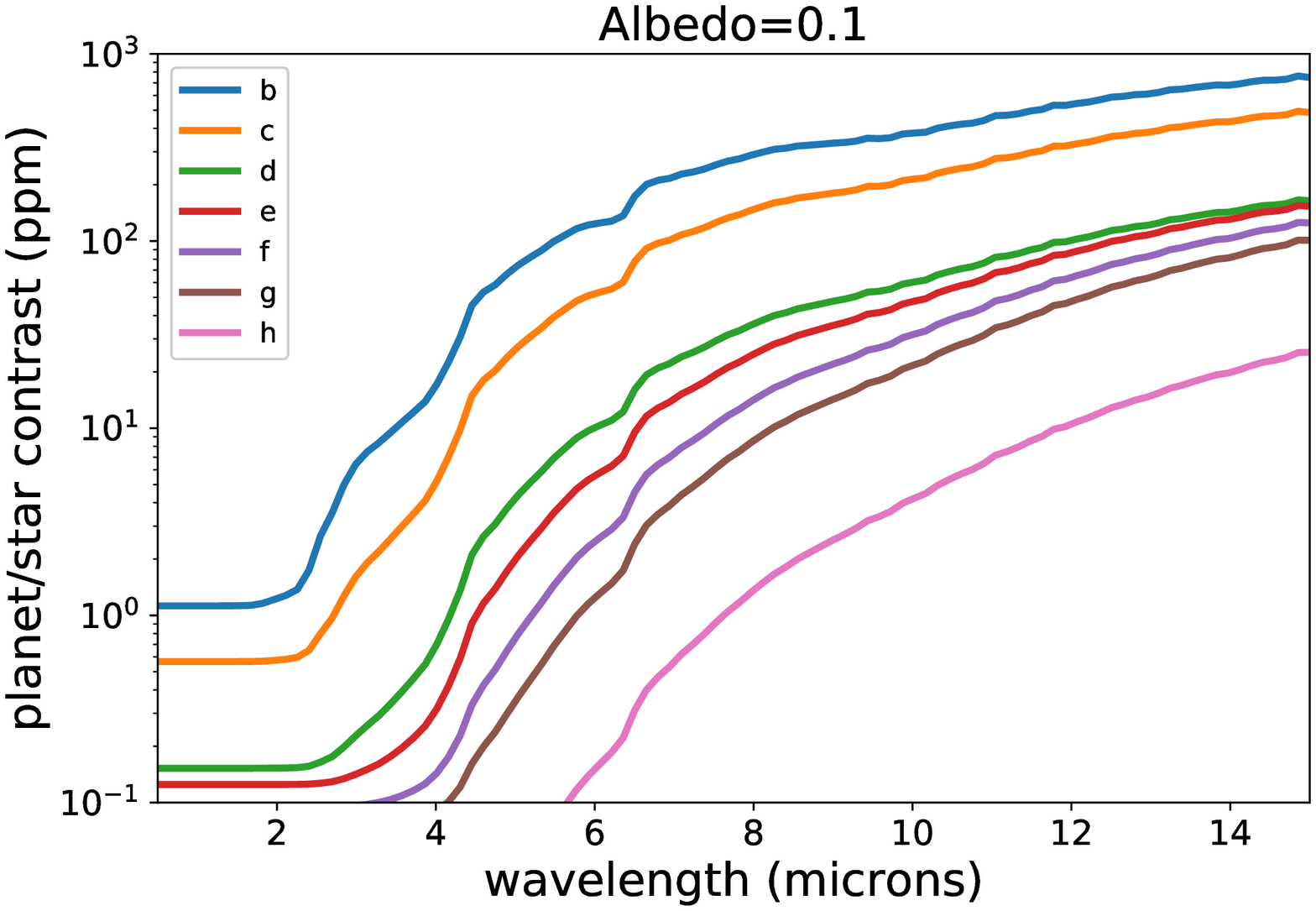} \\
    \includegraphics[width=8.5cm]{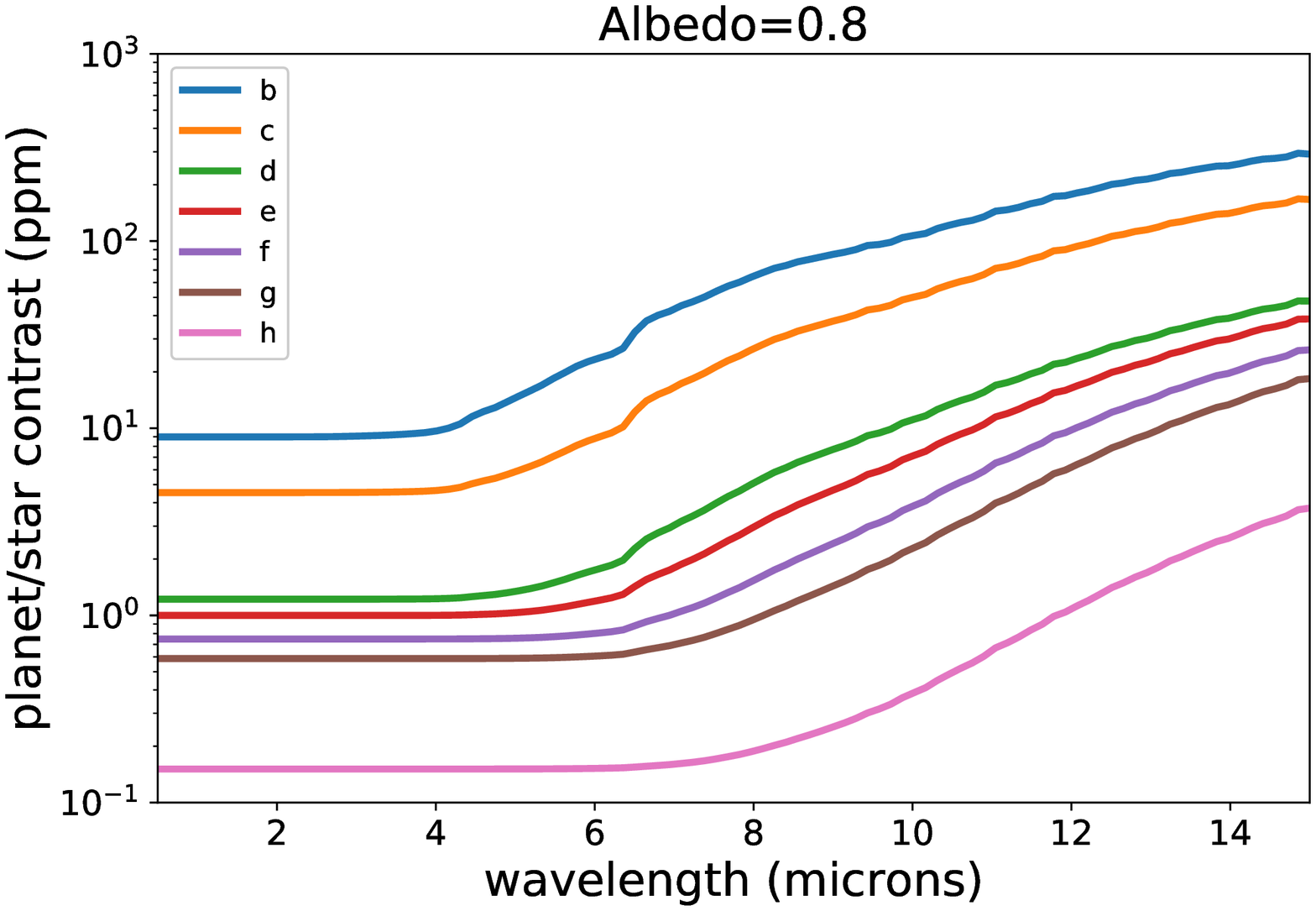}
  \end{center}
  \caption{Predicted eclipse depth for the known TRAPPIST-1 planets as
    a function of wavelength, assuming an albedo of 0.1 (top panel)
    and 0.8 (bottom panel).}
  \label{fig:eclipse}
\end{figure}

An important component of exoplanet phase variations is that
contributed by the infrared (IR) flux from the planet
\citep{selsis2011}. Using the calculated planetary equilibrium
temperatures provided by \citet{gillon2017a}, we estimated the IR flux
from each of the planets that contribute to the passband of various
instruments. Even the hottest of the planets, TRAPPIST-1b with a
temperature of $\sim$400~K, does not contribute significantly to the
integrated flux for {\it Kepler} or {\it TESS} passbands. However, the
eclipse depth will produce a stronger signature than that from phase
variations and will depend on the wavelength at which it was
observed. Shown in Figure~\ref{fig:eclipse} are the predicted eclipse
depths for each of the seven known TRAPPIST-1 planets as a function of
wavelength. These were calculated by integrating high resolution
spectra for the star and planet into 0.15~micron bins
\citep{baraffe2015}. The top and bottom panels assume extreme values
for the planetary albedos of 0.1 and 0.9, respectively. The models
also assume local Lambertian scattering for the reflected flux and
local radiative equilibrium for the thermal emission. As expected, the
high atmospheric absorption scenarios produce the most detectable
features at longer wavelengths. The IR signatures of the planets are
explored in the context of the GCM results provided in
Section~\ref{gcm}.

%%%%%%%%%%%%%%%%%%%%%%%%%%%%%%%%%%%%%%%%%%%%%%%%%%%%%%%%%%%%%%%%%%%%

\section{ROCKE-3D GCM Phase Curves}
\label{gcm}

%%%%%%%%%%%%%%%%%%%%%%%%%%%%%%%%%%%%%%%%%%%%%%%%%%%%%%%%%%%%%%%%%%%%

\subsection{Description of the Model}

We have employed the Resolving Orbital and Climate Keys of Earth and
Extraterrestrial Environments with Dynamics ({\sc ROCKE-3D}) GCM
\citep{way2017b} to simulate TRAPPIST-1e and TRAPPIST-1f
\citep{gillon2017a} atmospheric circulation using the updated planet
parameters from \citet{grimm2018} assuming a synchronous rotation and
aqua-planet configuration. The stellar spectrum of TRAPPIST-1 is
represented by a 2600~K BT Settl with [Fe/H] = 0
\citep{baraffe2015}. {\sc ROCKE-3D} was run at a $4^{\circ}\times
5^{\circ}$ ($46\times72$ coordinates) latitude-longitude resolution
with an altitude resolution of 40 layers up to 0.1~mb.

For TRAPPIST-1e, we have simulated both a 1~bar modern Earth-like
atmosphere mostly composed of N$_2$ and 400~ppm of CO$_2$, and a 1~bar
Archean-like atmosphere composed of N$_2$, 10,000~ppm of CO$_2$ and
2,000 ppm of CH$_4$ such as assumed in composition B of
\citet{charnay2013}. For TRAPPIST-1f, the modern Earth atmospheric
composition led to a fully frozen ocean from top to bottom and to the
crash of the simulation. As a result, only the Archean Earth
simulation from \citet{charnay2013} was simulated for TRAPPIST-1f.
Note that H$_2$O is treated as a variable specie, predicted by the
GCM.

\begin{table}
\centering
\caption{{\sc ROCKE-3D} bandpasses ($\mu$m)}
\label{tab:bands}
\begin{tabular}{l|cc}
 & modern Earth-like & Archean Earth-like \\ \cline{2-3} 
Short-wave & 0.200 - 0.320 & 0.200 - 0.385 \\
 & 0.320 - 5.050 & 0.385 - 0.500 \\
 & 0.505 - 0.690 & 0.500 - 0.690 \\
 & 0.690 - 0.770 & 0.690 - 0.870 \\
 & 0.770 - 0.890 & 0.870 - 0.900 \\
 & 0.890 - 0.980 & 0.900 - 1.080 \\
 & 0.980 - 1.040 & 1.080 - 1.120 \\
 & 1.040 - 1.100 & 1.120 - 1.160 \\
 & 1.100 - 1.160 & 1.160 - 1.200 \\
 & 1.160 - 1.240 & 1.200 - 1.300 \\
 & 1.240 - 1.340 & 1.300 - 1.340 \\
 & 1.340 - 1.420 & 1.340 - 1.420 \\
 & 1.420 - 1.520 & 1.420 - 1.460 \\
 & 1.520 - 1.620 & 1.460 - 1.520 \\
 & 1.620 - 1.800 & 1.520 - 1.560 \\
 & 1.800 - 1.960 & 1.560 - 1.620 \\
 & 1.960 - 2.380 & 1.620 - 1.680 \\
 & 2.380 - 2.950 & 1.680 - 1.800 \\
 & 2.950 - 4.100 & 1.800 - 1.940 \\
 & 4.100 - 4.500 & 1.940 - 2.000 \\
 & 4.500 - 20.00 & 2.000 - 2.140 \\
 & - & 2.140 - 2.500 \\
 & - & 2.500 - 2.650 \\
 & - & 2.650 - 2.850 \\
 & - & 2.850 - 3.150 \\
 & - & 3.150 - 3.600 \\
 & - & 3.600 - 4.100 \\
 & - & 4.100 - 4.600 \\
 & - & 4.600 - 20.00 \\ \hline
Long-wave & 3.333 - 4.950 & 3.339 - 4.149 \\
 & 4.950 - 6.667 & 4.149 - 4.566 \\
 & 6.667 - 7.519 & 4.566 - 7.092 \\
 & 7.519 - 8.130 & 7.092 - 7.634 \\
 & 8.130 - 8.929 & 7.634 - 8.333 \\
 & 8.929 - 10.10 & 8.333 - 8.929 \\
 & 10.10 - 12.50 & 8.929 - 10.10 \\
 & 12.50 - 13.33 & 10.10 - 12.50 \\
 & 13.33 - 16.95 & 13.33 - 16.95 \\
 & 16.95 - 18.18 & 12.50 - 18.18 \\
 & 18.18 - 25.00 & 18.18 - 25.00 \\
 & 25.00 - 10.00$\times10^{3}$ & 25.00 - 10.00$\times10^{3}$
\end{tabular}
\end{table}

We used SOCRATES\footnote{\tt https://code.metoffice.gov.uk/
  trac/socrates} radiation parameterization
\citep{edwards1996a,edwards1996b} to compute the radiative
transfer through the atmosphere as described by \citet{way2017b}. For
the modern Earth-like atmosphere, twelve bands in the longwave and
twenty-one bands in the shortwave have been used to build the GCM,
while twelve bands in the longwave and twenty-nine bands in the
shortwave have been used in the case of the Archean Earth-like
atmosphere (see Table~\ref{tab:bands} for the specific
wavebands). These spectral resolutions lead to fluxes accurate to
within several W/m$^2$ for planets orbiting an M dwarf such as
TRAPPIST-1. TRAPPIST-1e is assumed to be fully covered by a global
ocean \citep{unterborn2018a} with a horizontal resolution
$4^{\circ}\times 5^{\circ}$ and 13 vertical layers down to a 100~m
depth. The ocean includes ocean heat transport (OHT) with dynamic sea
ice parameterization \citep{way2017b} allowing fractional gridbox sea
ice cover. The simulations were run until the radiative balance at the
top of the atmosphere (TOA) was reached (e.g., a radiative imbalance
smaller than $\pm$0.2~Wm$^{-2}$).

%%%%%%%%%%%%%%%%%%%%%%%%%%%%%%%%%%%%%%%%%%%%%%%%%%%%%%%%%%%%%%%%%%%%

\subsection{Constructing the Phase Curve}

The {\sc ROCKE-3D} phase curves of the TRAPPIST-1 planets are modeled
as a sum of the outgoing shortwave radiation - a product of the
incident stellar radiation and planetary albedo - and the outgoing
longwave thermal radiation across the longitudes which are visible to
the observer at each point in phase. These radiation quantities are
delivered by {\sc ROCKE-3D} in static 2-dimensional grids spanning
latitude and longitude, having been averaged over 100 model years post
hydrological and radiative equilibrium. In order to represent the
photometric phase variations of the modeled TRAPPIST-1 system from the
perspective of an observer, we employ a moving window over the modeled
surface that integrates over the observable longitudes as a function
of phase and the given orbital dynamics, as described in the remainder
of this section. The angle definitions and nomenclature in the
following have been adapted from \citet{hu2015a}.

For a phase angle $\alpha$ defined such that $\alpha = 0$ at
occultation and $\alpha = \pm\pi$ at transit, the luminosity of the
hemisphere viewed by an observer as a function of phase can be
expressed simply as
\begin{equation}
    L(\alpha) = L_{SW}(\alpha) + L_{LW}(\alpha).
\end{equation}
Expanding the short-wave (SW) reflection component of the phase curve
gives
\begin{multline}\label{eq:lum}
    L_{SW}(\alpha) = \sum_{\phi =
      \phi_i(\alpha)}^{\phi_{j}(\alpha)}\sum_{\theta =
      -\frac{\pi}{2}}^{\frac{\pi}{2}} \left[ A_B(\xi(\phi), \theta)
      I(\xi(\phi), \theta) \right. \\ \times
      \left. \mathcal{A}(\xi(\phi), \theta) \cos{\theta}
      \cos{\phi}\right],
\end{multline}
where $A_B(\xi(\phi), \theta)$ is the Bond albedo of a specific
grid cell, and $I(\xi(\phi), \theta)$ is the incident stellar
radiation at the top of the atmosphere at the same grid cell, where
each grid cell is defined by its latitude $\theta$ and local longitude
$\xi$ (which itself depends on the longitude in the observer's frame
$\phi$ -- more on this later). The outgoing short-wave radiation is
then weighted by the grid cell area $\mathcal{A}(\xi(\phi), \theta)$
and the angle of its normal to the observer.

Adopting an inclination of 90$^{\circ}$, the range of visible
latitudes is constant with phase and defined to exist in $[-\pi/2,
  \pi/2]$ from the south pole to the north pole respectively. The
planetary longitude in the observer's frame is defined such that $\phi
= -\pi /2$ at the west terminator, $\phi = \pi /2$ at the east
terminator, and $\phi = 0$ in the direction of the observer.

For the reflection component of the phase curve, the relevant range of
longitudes [$\phi_i, \phi_j$] is that which appears illuminated at a
given phase. The western-most illuminated longitude $\phi_i$ and
eastern-most illuminated longitude $\phi_j$ in the frame of the
observer can be described as a function of phase as
\begin{equation}\label{eq:phi_ref}
\phi_x(\alpha) =
\begin{cases}
        \gamma \alpha - \pi / 2 & x=i \\
        -(\gamma - 1) \alpha + \pi / 2 & x=j\\
\end{cases}
\end{equation}
where 
\begin{equation*}
    \gamma = 
    \begin{cases}
        0 & -\pi < \alpha \le 0 \\
        1 & \ \ \ 0 \le \alpha < \pi
    \end{cases}.
\end{equation*}
These longitudes are then translated to the corresponding columns of
the GCM grid (i.e. the ``local longitudes'' in the planetary frame,
for which we use the symbol $\xi$) by considering the local longitude
facing the observer at observation start, the planet's rotation
frequency and orbital period, and the phase angles elapsed over the
observation:
\begin{equation}\label{eq:loc_long_x}
    \xi(\phi_x, \alpha) = \phi_x(\alpha) + \xi_{\hat{o}}(\alpha),
\end{equation}
where $\xi_{\hat{o}}$ is the local longitude in the direction of the
observer at a given phase. For a planet on a prograde orbit, this is
equal to
\begin{equation}\label{eq:xi}
    \xi_{\hat{o}}(\alpha) = \left[ \xi_0 - P f_\mathrm{rot}(\alpha -
      \alpha_0) \right] \mod{2\pi}
\end{equation}
where $\alpha_0$ is the phase angle at the start of the observation,
$\xi_0$ is the local longitude in the direction of the observer at the
start of the observation, $P$ is the orbital period in days, and
$f_\mathrm{rot}$ is the rotation frequency of the planet in
days$^{-1}$. Assuming a synchronously rotating planet, as we do for
TRAPPIST-1e and f, $Pf_\mathrm{rot} = 1$.

The long-wave component of the phase curve extracted from {\sc
  ROCKE-3D} at a given phase angle can be similarly expressed as
\begin{multline}
    L_{LW}(\alpha) = \sum_{\phi =
      -\frac{\pi}{2}}^{\frac{\pi}{2}}\sum_{\theta =
      -\frac{\pi}{2}}^{\frac{\pi}{2}} \left[F_T(\xi(\phi), \theta)
      \right. \\ \left. \times \mathcal{A}(\xi(\phi), \theta)
      \cos{\theta} \cos{\phi}\right]
\end{multline}
where $F_T(\xi(\phi), \theta)$ is the outgoing net thermal flux at the
top of the atmosphere at a specific grid cell. Because the entire
surface emits long-wave radiation throughout its orbit, the western
and eastern-most radiating longitudes in the frame of the observer are
constant in phase for the thermal component of the phase
curve. Translating these longitudes to the local longitudes can then
be done using Equation~\ref{eq:loc_long_x}, redefining the first term
$\phi_x$ as
\begin{equation}\label{eq:phi_therm}
  \phi_x = 
  \begin{cases}
    -\pi / 2 & x=i \\
    \pi / 2 & x=j\\
  \end{cases}.
\end{equation}

\begin{figure*}
  \begin{center}
    \includegraphics[clip,width=10.0cm]{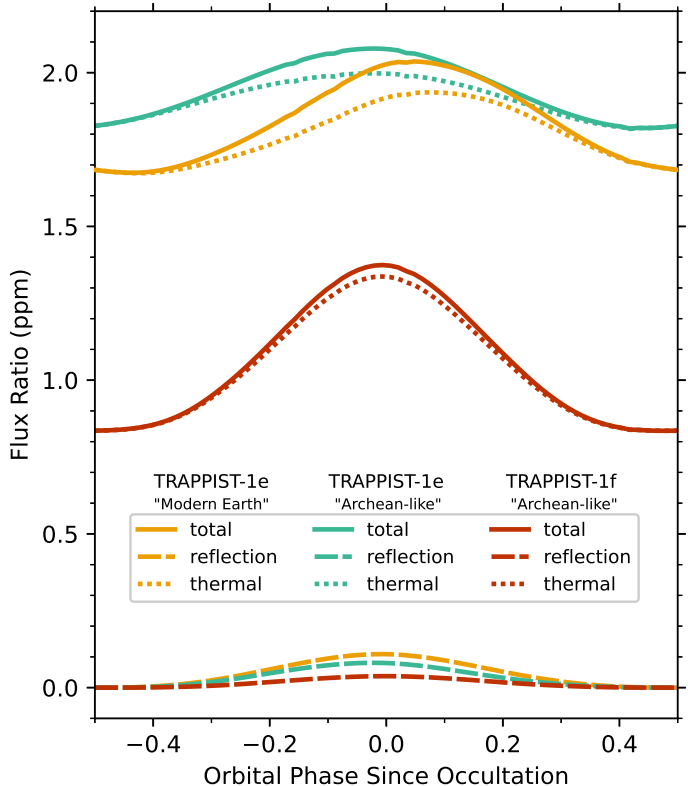} \\
  \end{center}
  \caption{Phase curves of the ROCKE-3D GCM simulations of TRAPPIST-1e
    with a ``modern Earth'' atmospheric configuration of 1 bar of
    N$_2$ and 400 ppm of CO$_2$ (yellow), an ``Archean-like'' model of
    TRAPPIST-1e with 1bar of N$_2$, 10,000 ppm of CO$_2$ and 2,000 ppm
    of CH$_4$ from \cite{charnay2013} (teal), and a model of
    TRAPPIST-1f with the same Archean-like atmospheric composition
    (red). Dashed lines show the reflection components of the phase
    curves integrated over bandpasses spanning 0.2--20 microns, while
    the dash-dotted lines show the long-wave thermal components over a
    wavelength range spanning 3.33--10,000 microns (see
    Table~\ref{tab:bands} for the specific spectral resolution). Solid
    lines show the sum of the reflection and thermal components,
    i.e. the total outgoing radiation from the top of the atmosphere
    over a wavelength range of 0.2--10,000 microns. Each curve is
    phased on the orbital period of the respective planet. Note: In a
    true photometric observation the transit and occultation events
    would be present in the phase curves, but are not included here.}
  \label{fig:rocke3d_phasecurves}
\end{figure*}

Because the model data are given in a relatively low resolution of
latitude and longitude, the longitude bounds in the discrete
summations described in this section likely fall somewhere between
grid lines, in which case only a fraction of the longitude will be
visible to the observer. To correct for this, we create a linearly
interpolated function of longitudinal surface luminosity at each point
in phase which we use to perform the discrete sums. These surface
luminosities are then divided by the stellar luminosity emitted by one
hemisphere (integrated over the corresponding short and long-wave
bands) to produce flux ratios. Note that the isotropic approximation
of the surface luminosity (see Equation~\ref{eq:lum}) does not take
into account atmospheric absorption or back-scattering effects.

%%%%%%%%%%%%%%%%%%%%%%%%%%%%%%%%%%%%%%%%%%%%%%%%%%%%%%%%%%%%%%%%%%%%

\subsection{Model Phase Curve Discussion}
\label{mpcd}

\begin{figure*}
  \begin{center}
    \includegraphics[clip,width=17.0cm]{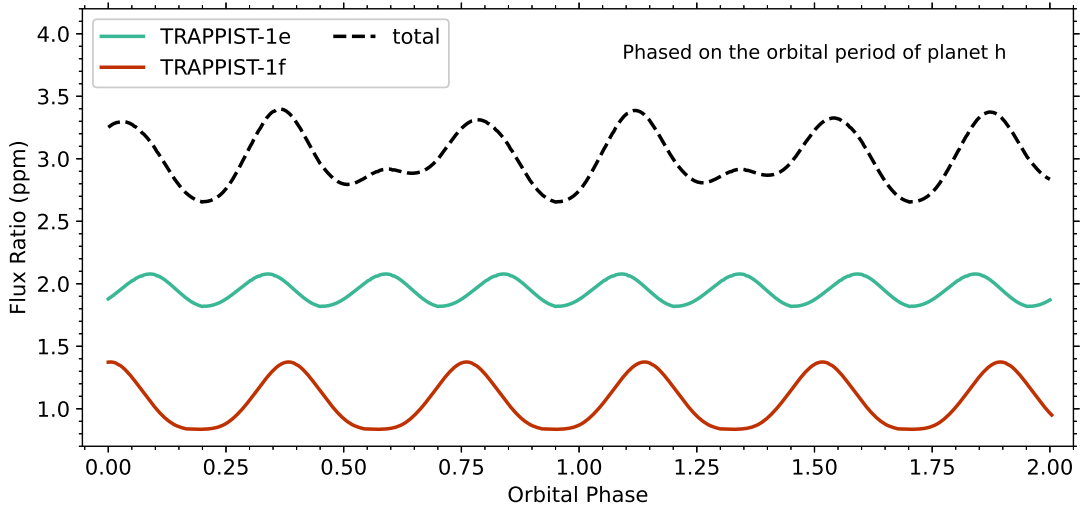}
  \end{center}
  \caption{Phase curves of the ROCKE-3D generated models of
    TRAPPIST-1e (teal) \& f (red), both with the atmospheric models of
    composition B in \cite{charnay2013}. Both curves have been phased
    on the orbital period of TRAPPIST-1h to showcase how the phase
    curves of the two planets can interfere (dashed line). Note: In a
    true photometric observation the transit and occultation events
    would be present in the phase curves, but are not included here.}
  \label{fig:rocke3d_phased-on-h}
\end{figure*}

Figure~\ref{fig:rocke3d_phasecurves} shows the phase curves extracted
from the ROCKE-3D GCM simulations for TRAPPIST-1e with the ``modern
Earth'' atmosphere and for TRAPPIST-1e and TRAPPIST-1f with an
``Archean-like'' atmosphere in both the wide shortwave bandpass
spanning 0.2--20~$\mu$m, and the wide longwave bandpass spanning
3.33--10,000~$\mu$m. The combination of the planet e and f phase
variations, phased on the orbit of planet h, are shown in
Figure~\ref{fig:rocke3d_phased-on-h}. The amplitudes of the phase
variations and the effect of their combination are comparable to those
predicted in Figure~\ref{fig:combined} and Table~\ref{tab:flux}, where
recall that the albedo was set to unity.

\begin{figure*}
  \begin{center}
    \begin{tabular}{cc}
      \includegraphics[clip,angle=90,width=8.5cm]{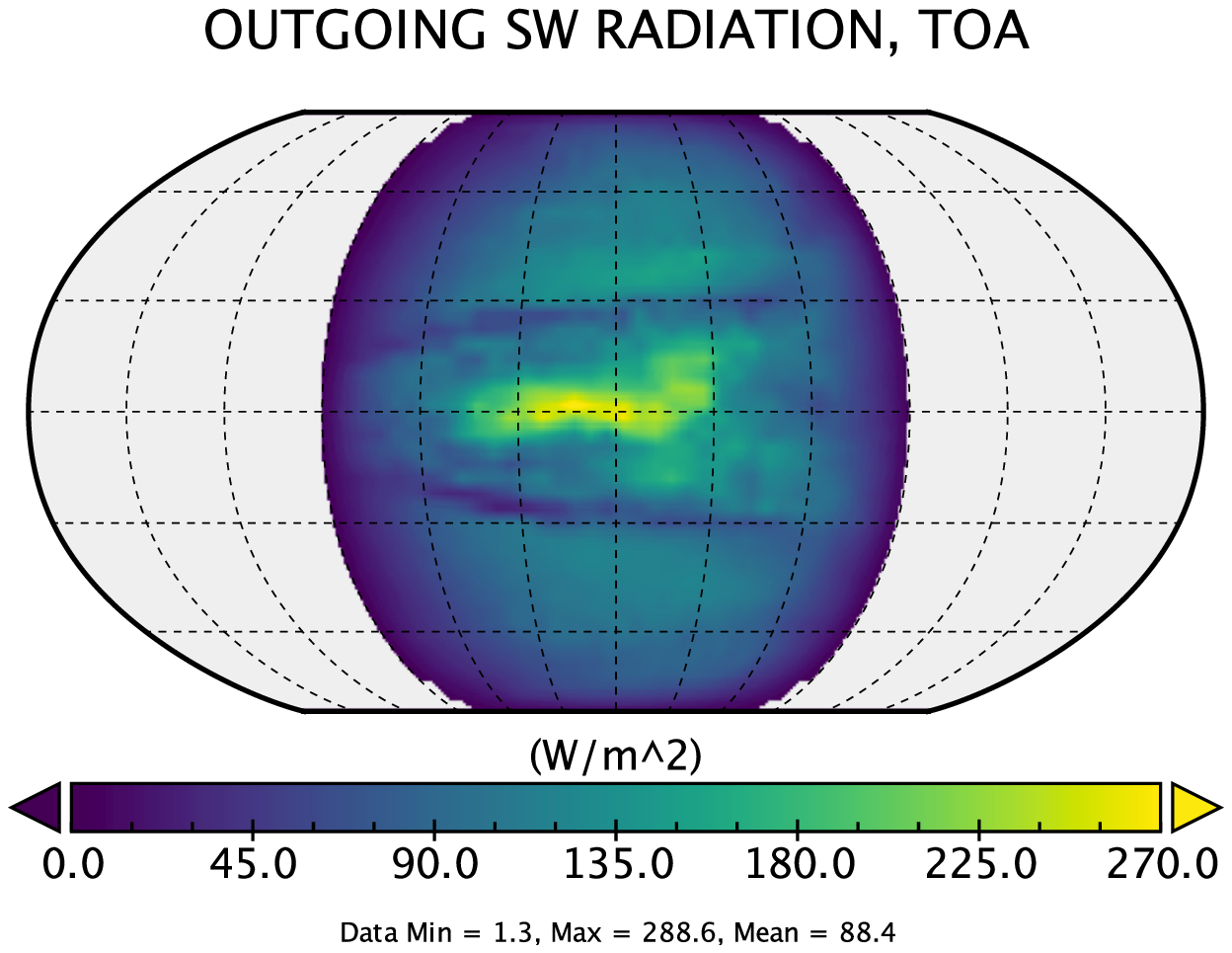} &
      \includegraphics[clip,angle=90,width=8.5cm]{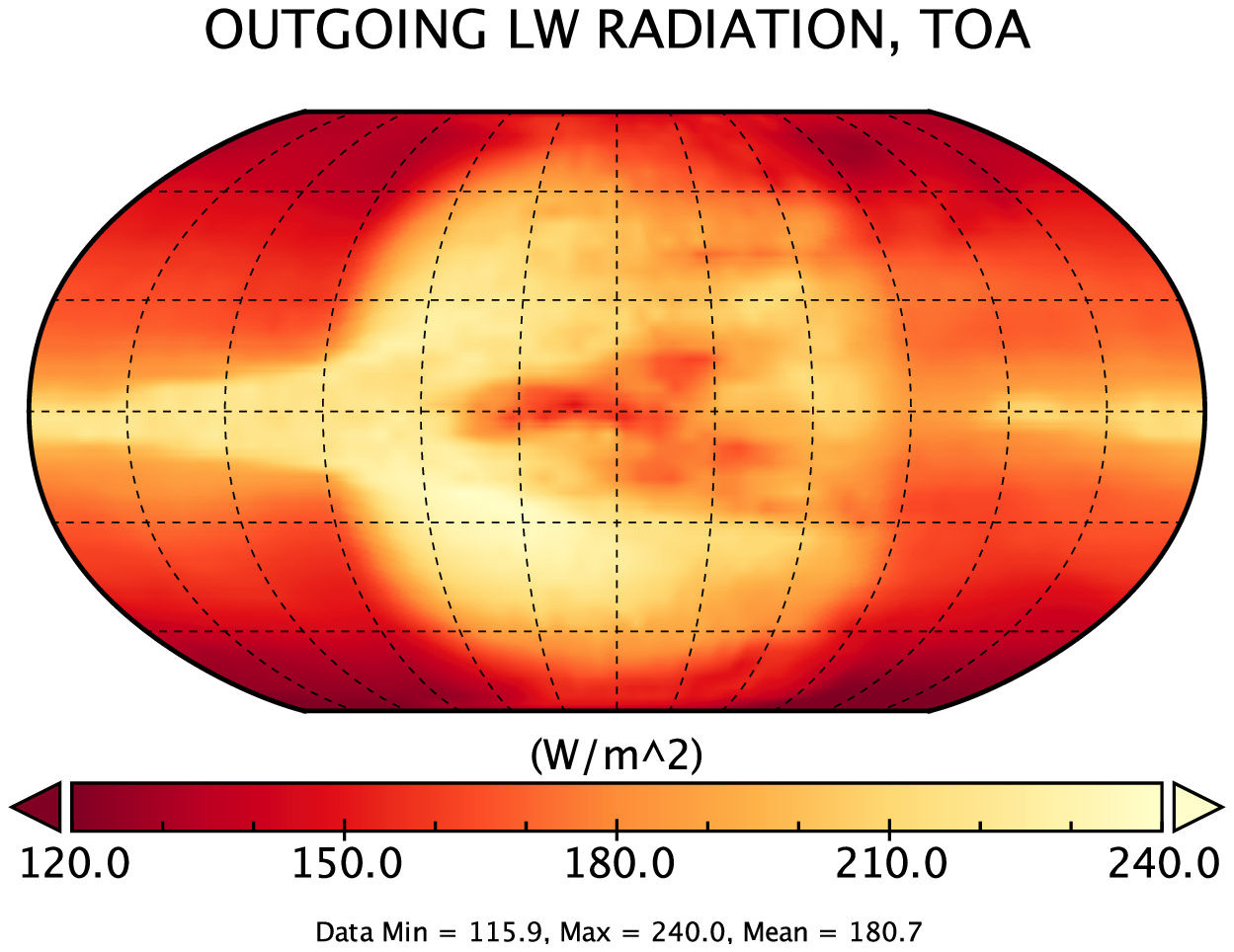} \\
      \includegraphics[clip,angle=90,width=8.5cm]{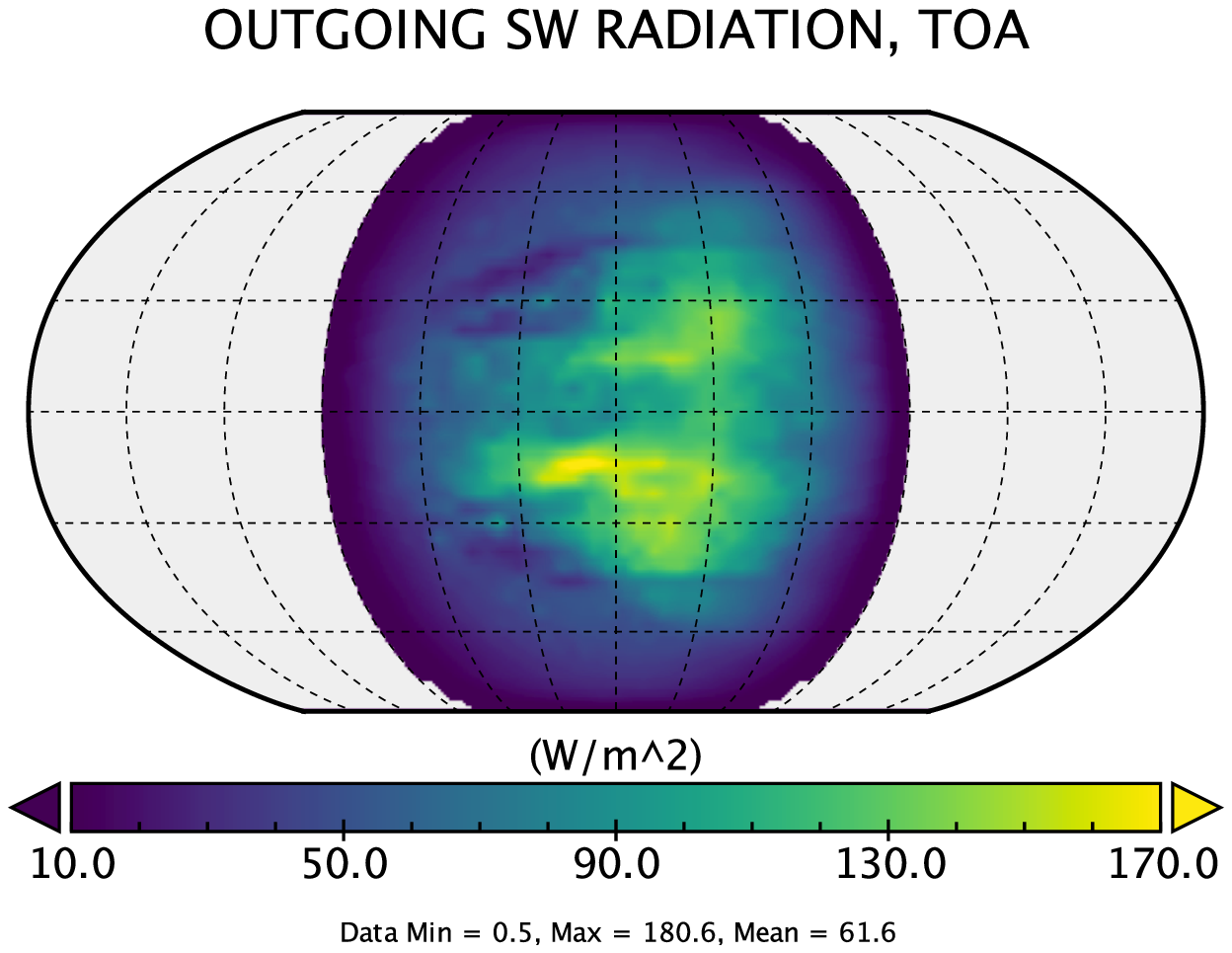} &
      \includegraphics[clip,angle=90,width=8.5cm]{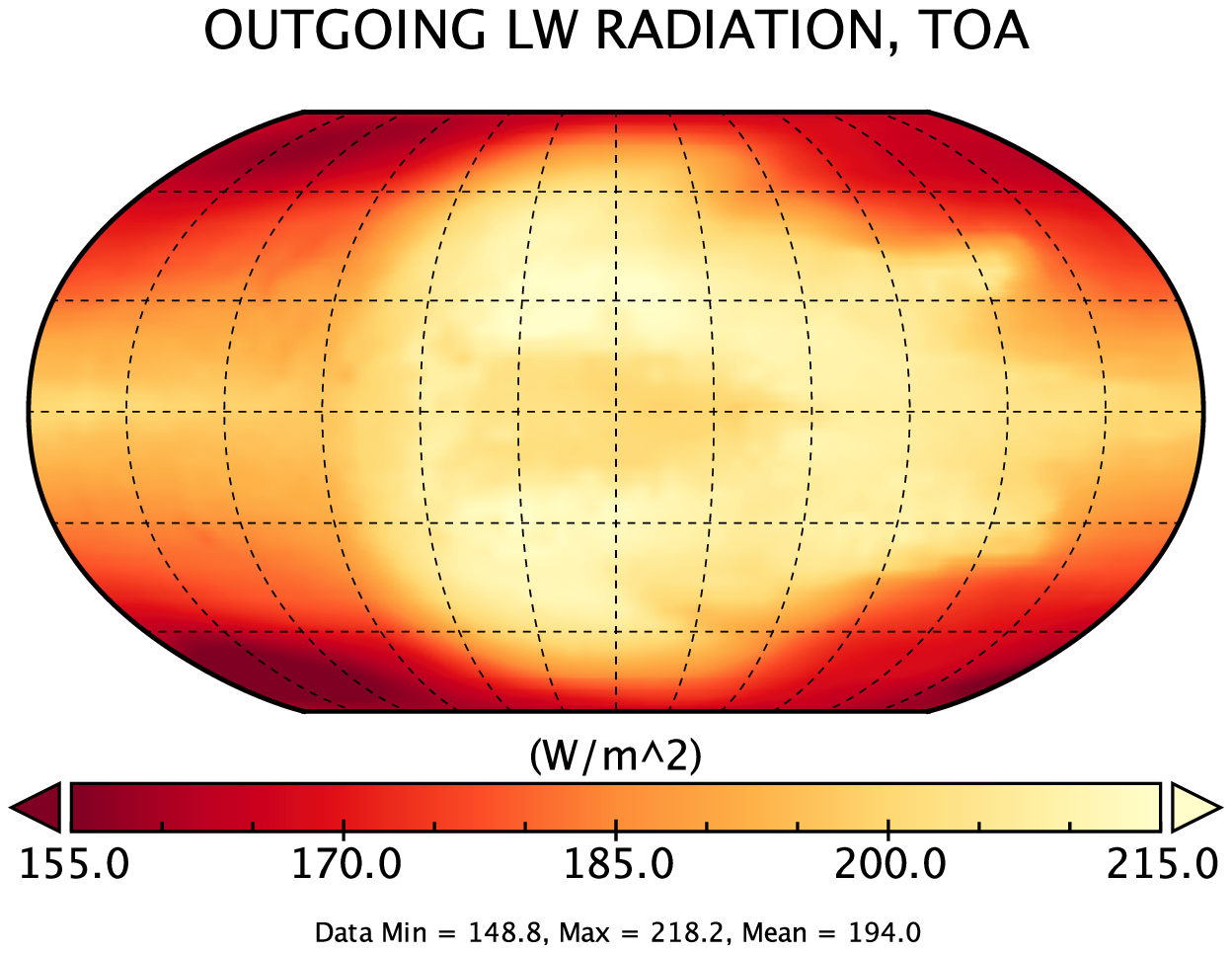} \\
      \includegraphics[clip,angle=90,width=8.5cm]{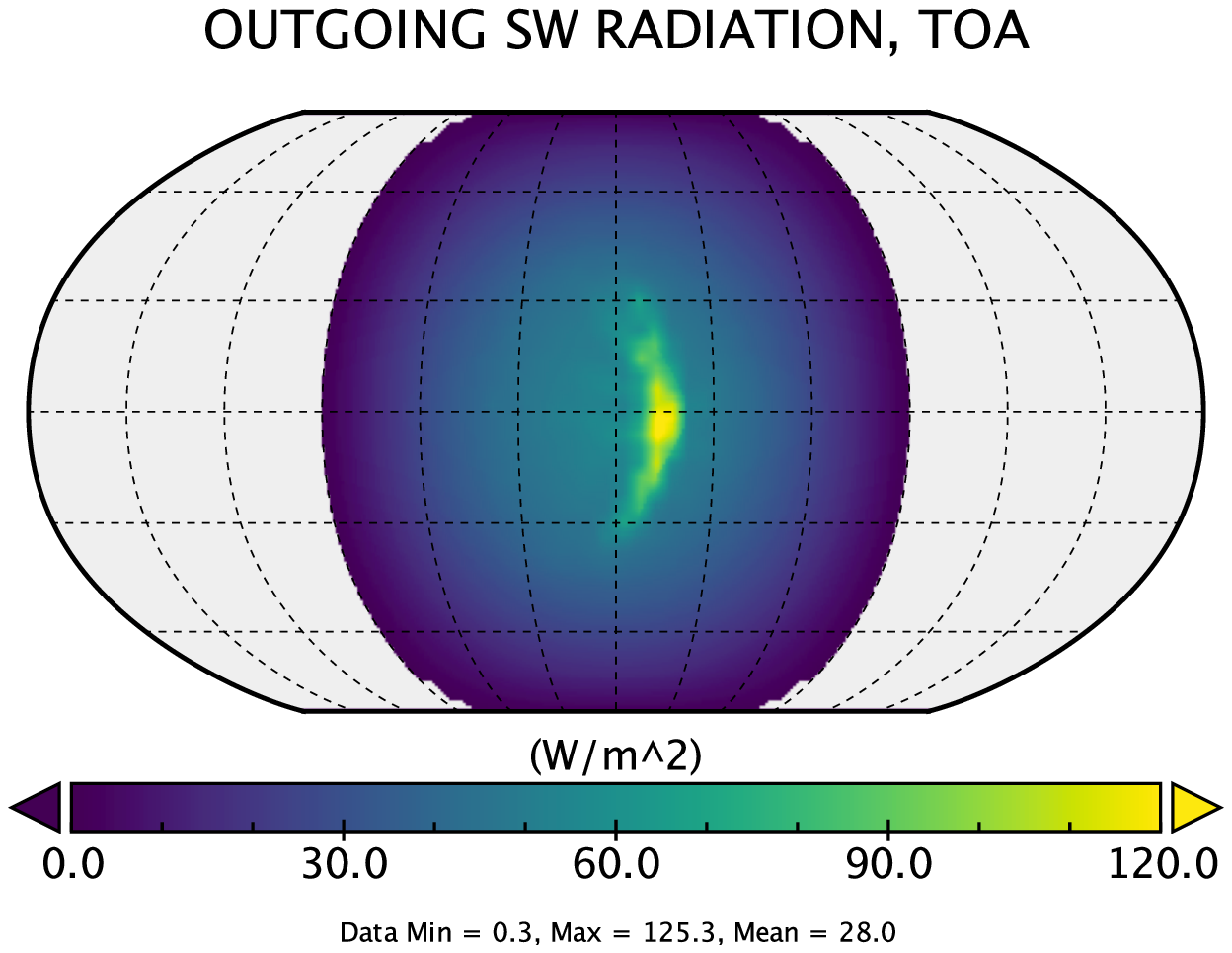} &
      \includegraphics[clip,angle=90,width=8.5cm]{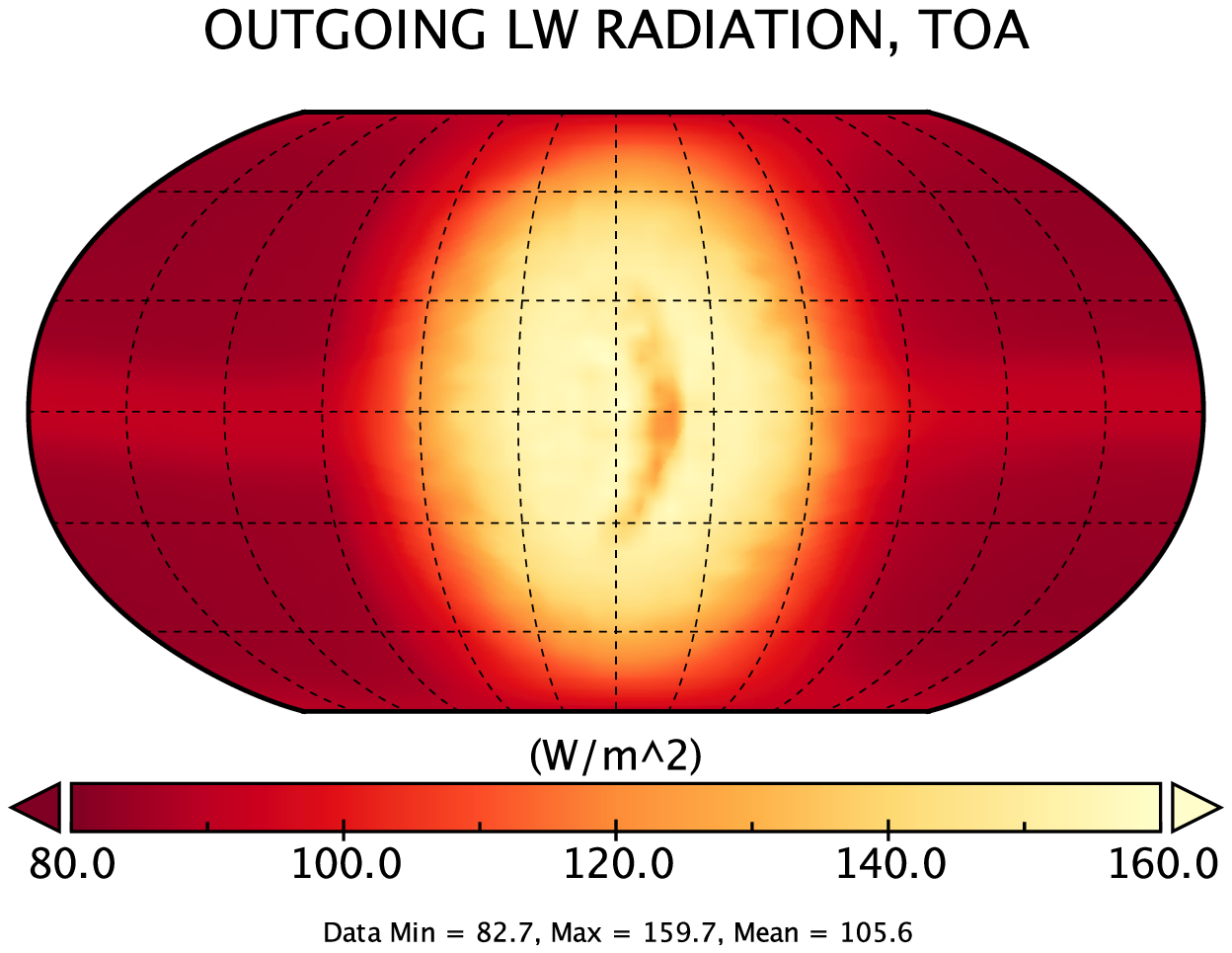} \\
    \end{tabular}
  \end{center}
  \caption{Robinson projections of the outgoing short-wave (left
    column) and long-wave (right column) radiation from the top of the
    simulated atmospheres of TRAPPIST-1e \& f created by the ROCKE-3D
    GCM. Each row displays the results of the three simulations
    explored in this study: the TRAPPIST-1e ``modern Earth''
    atmosphere (top), the TRAPPIST-1e ``Archean-like'' atmosphere
    (middle), and the TRAPPIST-1f ``Archean-like'' atmosphere
    (bottom). The sub-stellar point is at the center of each map,
    where the gray areas in the left column show the sides of the
    synchronously rotating planets which do not receive stellar
    radiation. Grid lines represent 30$^{\circ}$ in longitude and
    latitude. Note that the color in each map is scaled differently to
    highlight the unique radiation patterns of each surface.}
  \label{fig:rocke3d_maps}
\end{figure*}

The cause for the differences between the TRAPPIST-1e ``modern Earth''
phase curves and the ``Archean'' model phase curves, both in terms of
the shift in the peak amplitude and the relative asymmetry, lies in
the differences between the distribution of the outgoing radiation
across their surfaces. Maps of the outgoing radiation for each model
can be examined in the Robinson projections shown in
Figure~\ref{fig:rocke3d_maps}. The rows of
Figure~\ref{fig:rocke3d_maps} display the results for the TRAPPIST-1e
modern Earth-like atmosphere (top), the TRAPPIST-1e Archean-like
atmosphere (middle), and the TRAPPIST-1f Archean-like atmosphere
(bottom). The columns are the outgoing short-wave (left) and long-wave
(right) radiation from the top of the simulated atmospheres. The
sub-stellar point is at the center of each map, where the gray areas
in the left column show the sides of the synchronously rotating
planets which do not receive stellar radiation.  For the long-wave
radiation, the westerly-dominant asymmetry about the substellar
longitude seen in the ``Modern Earth'' TRAPPIST-1e map results in a
shift of the phase curve maximum of 17$^\circ$ post-occultation, which
can be seen in Figure~\ref{fig:rocke3d_phasecurves}. Interestingly,
the thermal radiation in the Archean-like model of the same planet
extends in the opposite (eastern) direction compared to the model with
the more modern Earth-like atmospheric composition, which results in a
shift of the phase curve maximum of 7.4$^\circ$ prior to
occultation. Likewise, the case of the TRAPPIST-1f long-wave radiation
is relatively symmetric around the sub-stellar point (with a slight
westward shift of 2.5$^\circ$), explaining the nearly symmetric phase
curve calculated for TRAPPIST-1f shown in
Figure~\ref{fig:rocke3d_phasecurves}. It is worth noting that the
perturbation of Keplerian orbits detected in the form of TTVs
\citep{grimm2018,agol2020b} will also affect the shift in the peak
phase amplitudes. These TTV effects are relatively small, varying the
true longitude variations of the planetary orbits by less than
$\pm0.4$\% of the orbit, compared with with eclipse duration of
$0.6$\% of the orbit.

\begin{figure*}
  \begin{center}
    \begin{tabular}{cc}
      \includegraphics[clip,width=9cm]{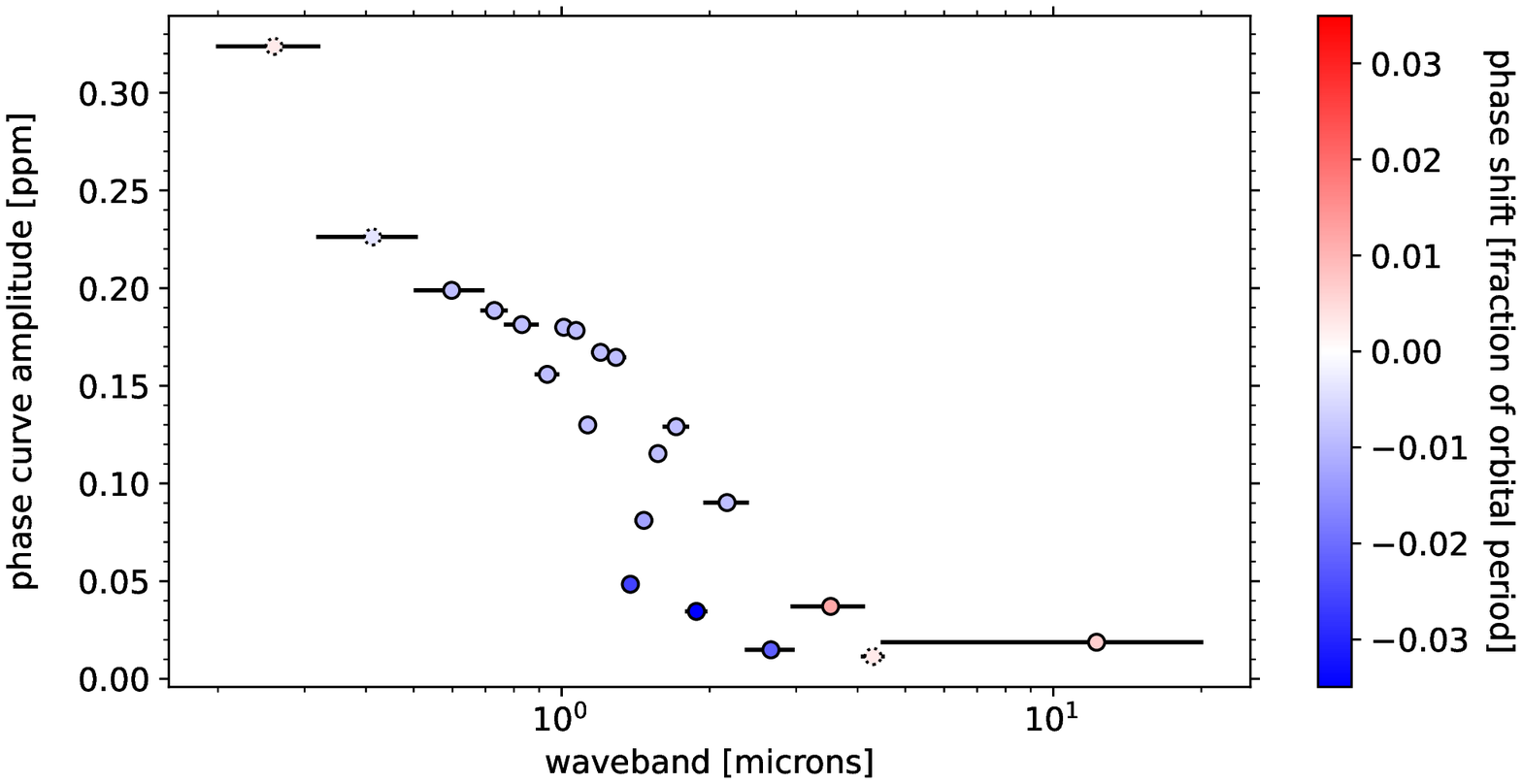} &
      \includegraphics[clip,width=9cm]{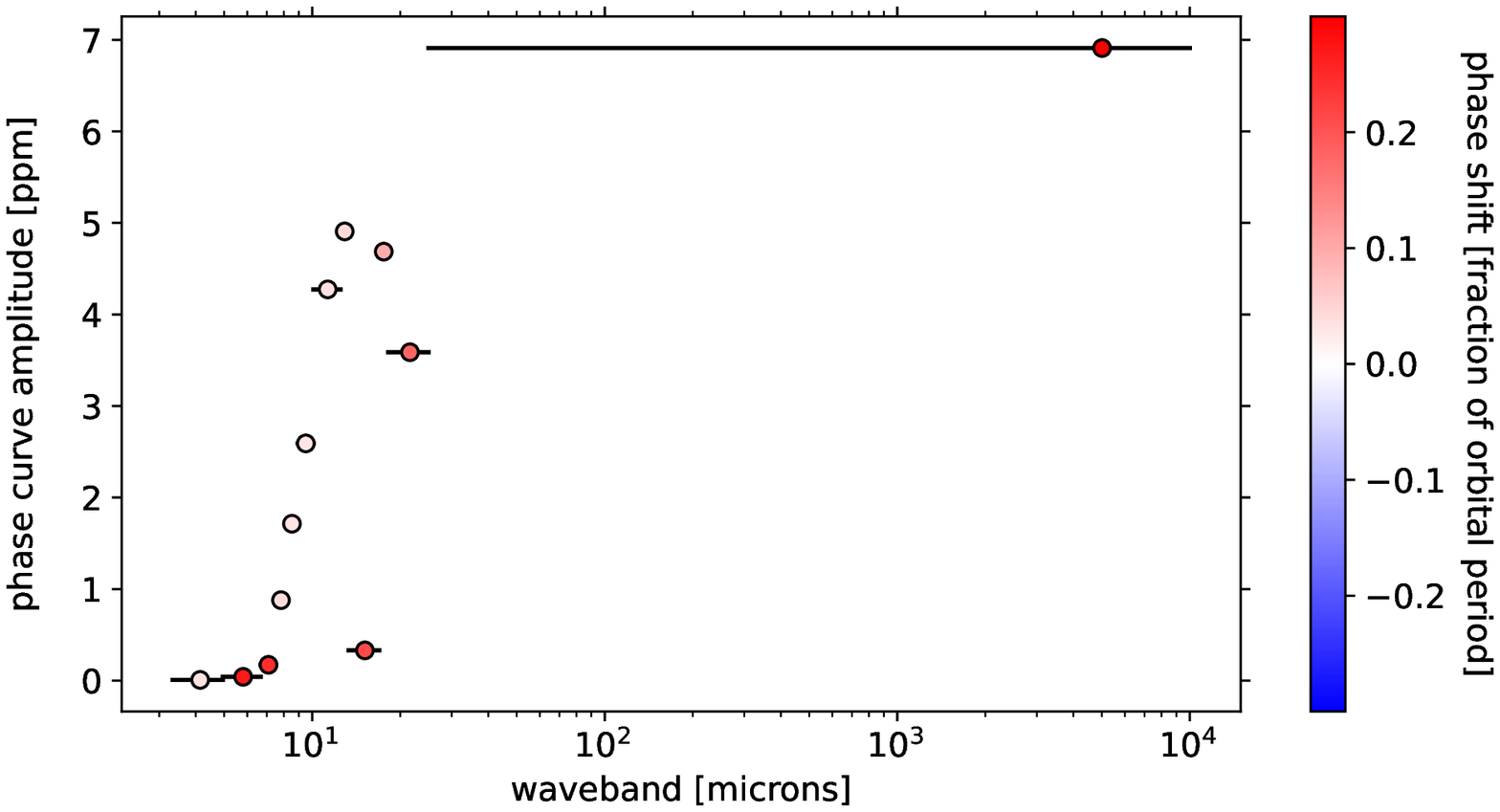} \\
      \includegraphics[clip,width=9cm]{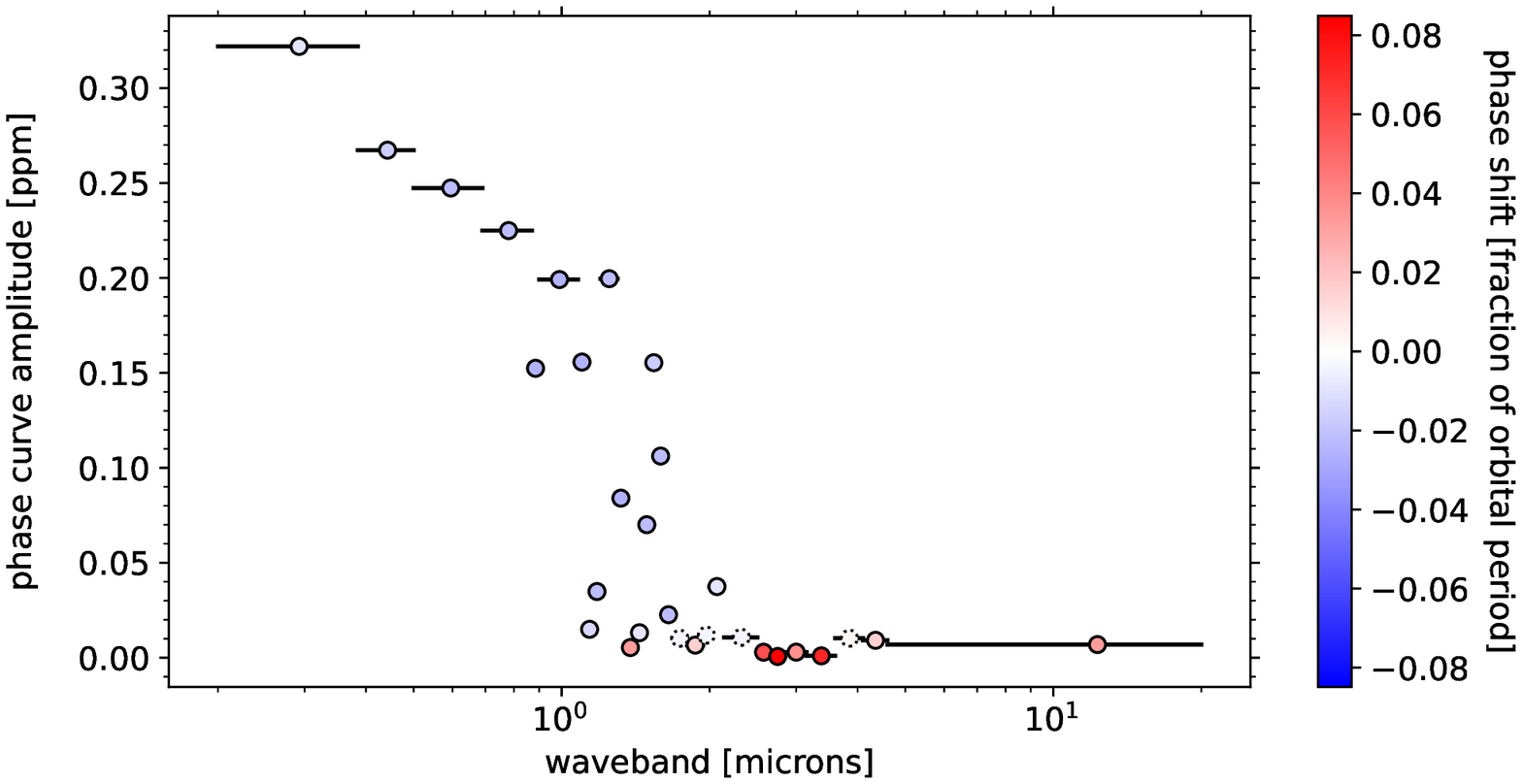} &
      \includegraphics[clip,width=9cm]{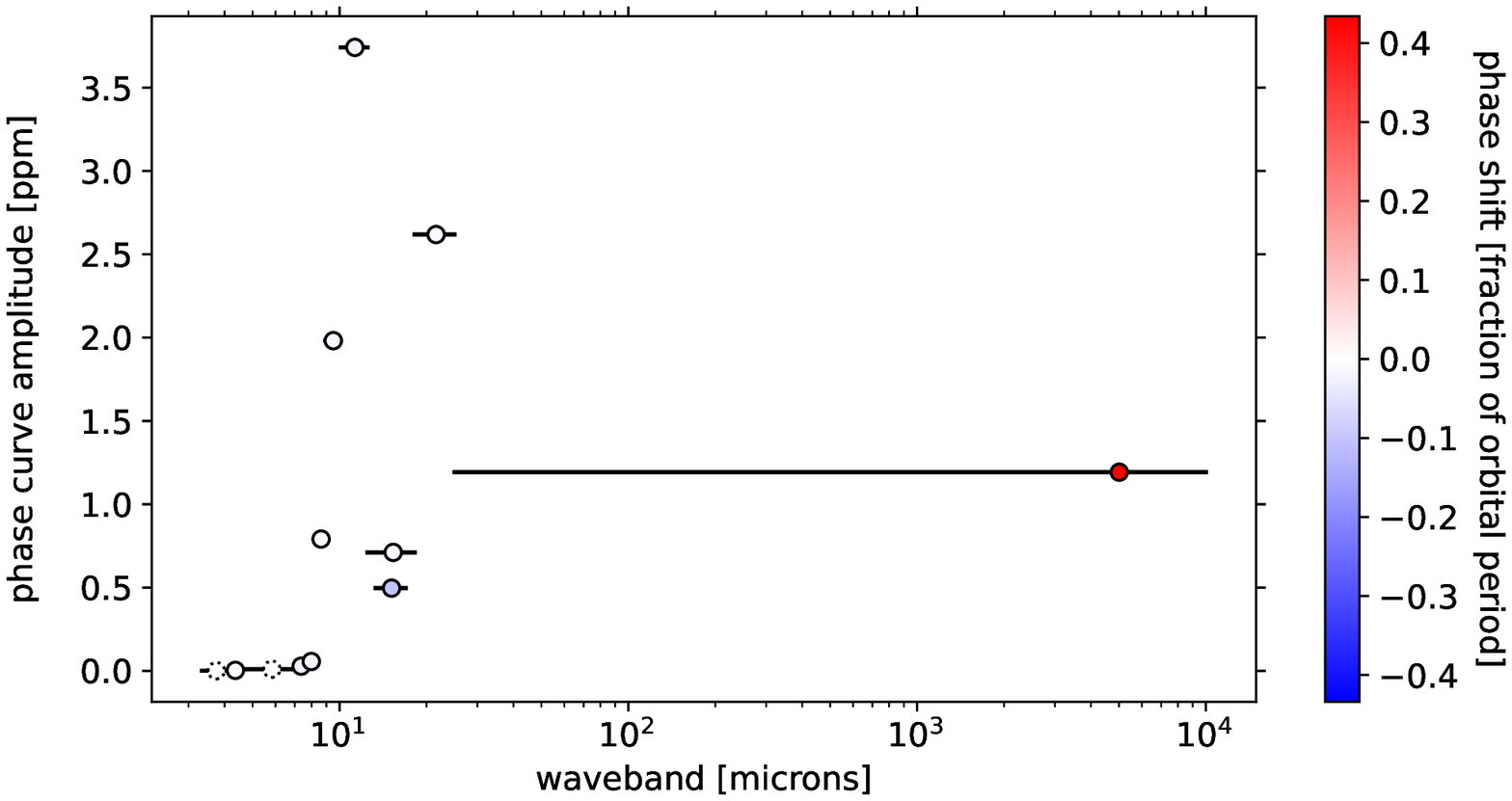} \\
      \includegraphics[clip,width=9cm]{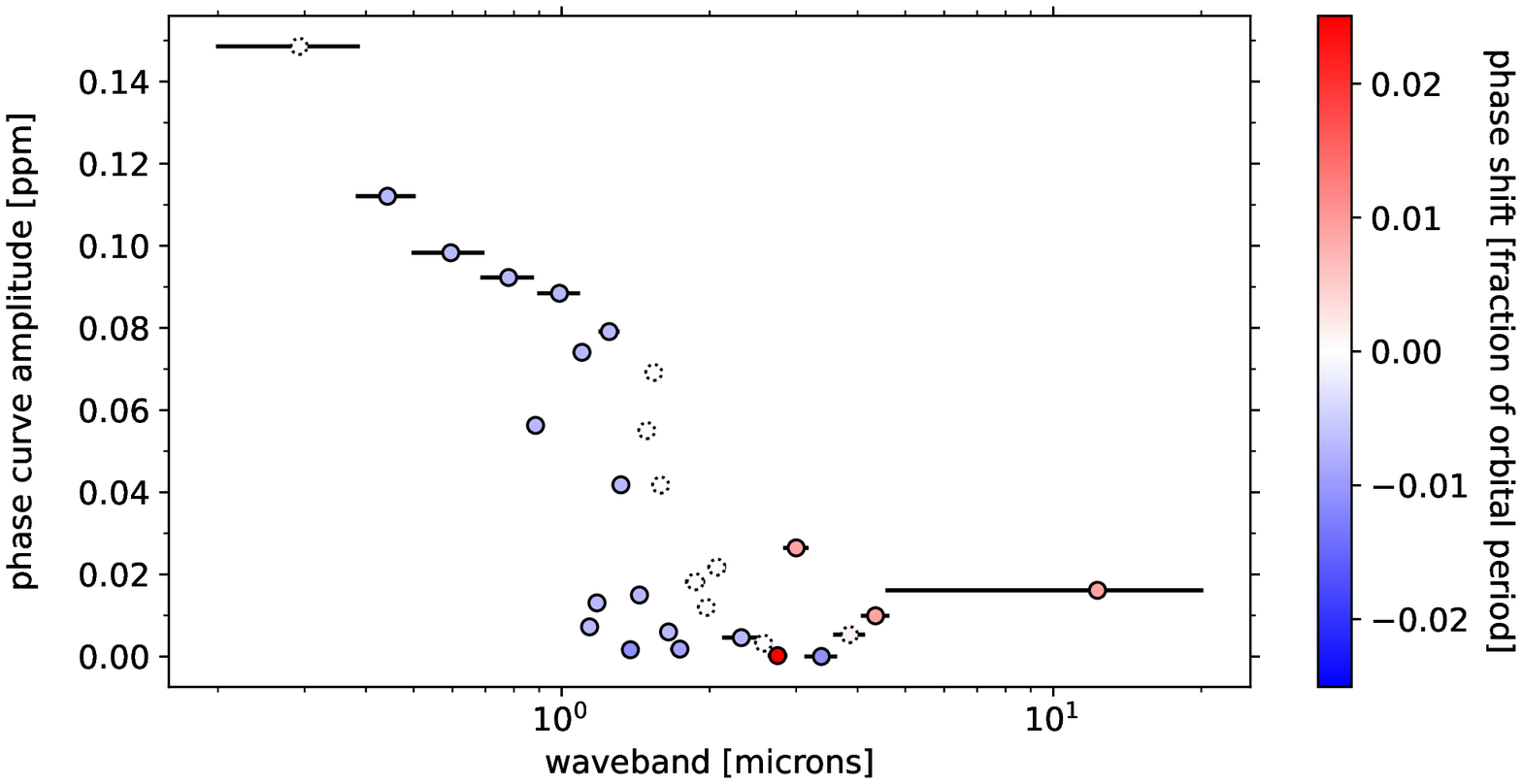} &
      \includegraphics[clip,width=9cm]{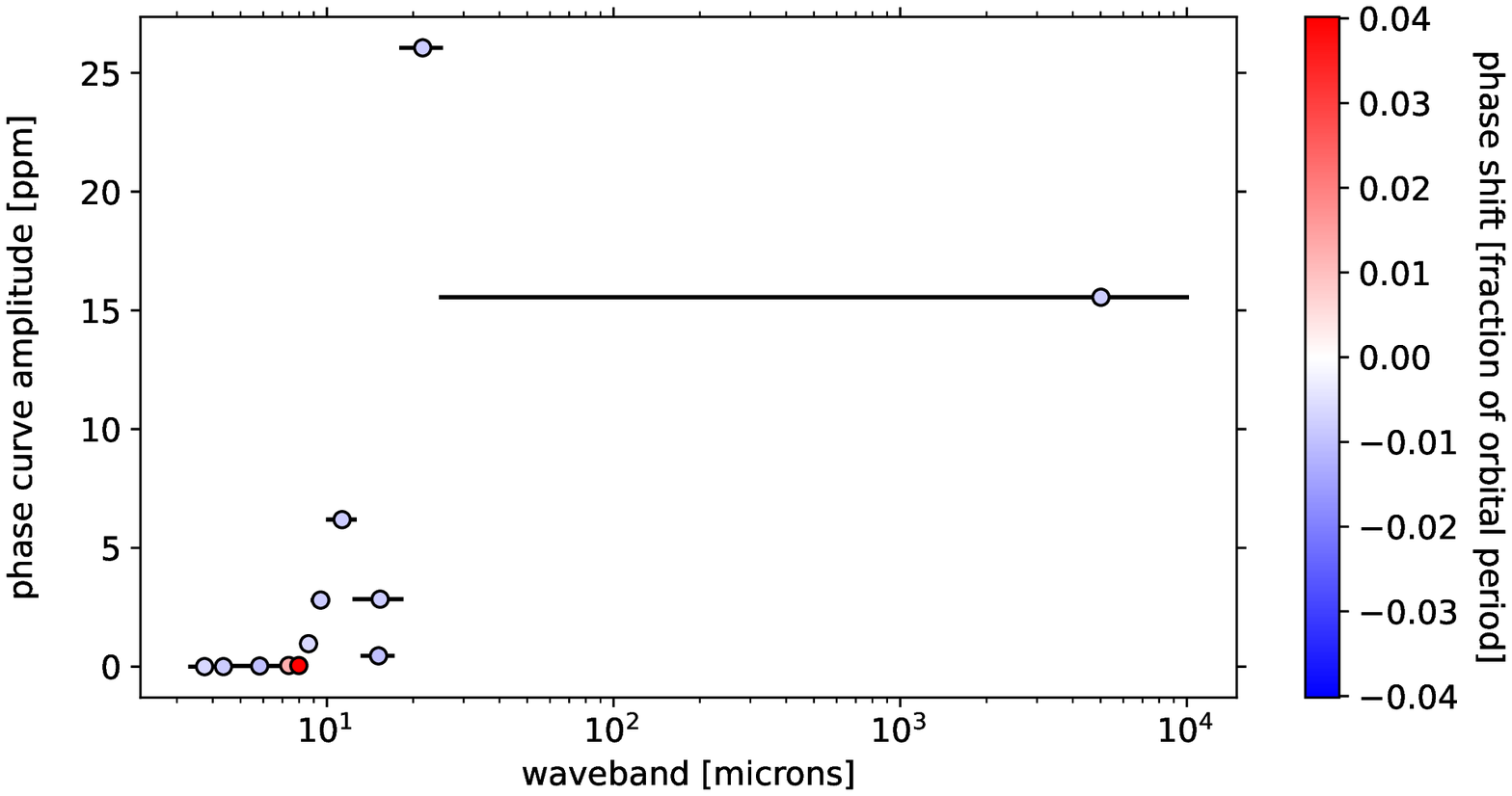} \\
    \end{tabular}
  \end{center}
  \caption{Peak-to-trough phase curve amplitudes for the separate
    wavebands covered in the ROCKE-3D GCM. Marker locations on the
    x-axis indicate the central wavelength in each band. Marker colors
    represent a shift in the phase curve maximum from superior
    conjunction, where markers with dotted outlines have peak
    amplitudes that are hidden behind the star during
    occultation. Redder points indicate a post-occultation maximum
    while bluer points indicate a pre-occultation maximum. These
    shifts are caused by a concentration of outgoing radiation on
    either the eastern hemisphere (blue) or western hemisphere (red)
    relative to the substellar point. The lefthand column shows the
    reflected shortwave radiation, while the righthand column shows
    the outgoing thermal radiation (see Table~\ref{tab:bands} for the
    specific bandpasses). The top row shows the modern Earth-like
    model for TRAPPIST-1e, middle row the Archean Earth-like model for
    TRAPPIST-1e, and bottom row the Archean Earth-like model for
    TRAPPIST-1f.}
  \label{fig:waveband_ppm}
\end{figure*}

Figure~\ref{fig:waveband_ppm} shows the phase curve amplitudes and
shifts of their maxima in each of the wavebands listed in
Table~\ref{tab:bands}. Here we can see that the phase shifts of the
phase curve maxima and the peak-to-trough amplitudes can vary quite
considerably with wavelength. Comparing the phase curves in the
shortwave bandpasses (left column) to those in the longwave bandpasses
(right column) shows that the shortwave phase curve amplitudes are
less than 1 ppm for each TRAPPIST model, and are consistently smaller
than the longwave phase curve amplitudes, which can be as high as 25
ppm for the TRAPPIST-1f model (owing to the relatively low thermal
redistribution to the nightside). The most promising signal from the
models comes from the phase curve of the modern Earth-like TRAPPIST-1e
in the farthest infrared waveband, which produces a peak-to-trough
amplitude of nearly 7 ppm. Though it has a lower amplitude than the 25
ppm signal produced by the ``Archean'' TRAPPIST-1f model, the ``modern
Earth'' model of TRAPPIST-1e radiates more photons overall, and,
having a shorter orbital period, can be observed over more orbits in
the same amount of time. Interestingly, the phase curve maxima in the
same FIR band for the TRAPPIST-1e simulations occur 0.30 orbital
periods post-occultation for the ``modern Earth'' model and 0.43
orbital periods post-occultation for the ``Archean Earth'' model,
which are caused by a concentration of outgoing radiation from the
nightsides of these simulated planets.

%%%%%%%%%%%%%%%%%%%%%%%%%%%%%%%%%%%%%%%%%%%%%%%%%%%%%%%%%%%%%%%%%%%%

\section{Discussion}
\label{discussion}

Evaluating the nature of of the TRAPPIST-1 planets and their
atmospheres remains a continuing focus for much of the exoplanet
community \citep{turbet2020c}. Numerous groups have been formed to
study the TRAPPIST-1 atmospheres, including an advocation for robust
comparison of atmospheric models \citep{fauchez2020b}. Studies include
predictions of potential biosignatures \citep{hu2020} and
recommendations for their interpretation
\citep{fujii2018,schwieterman2018}, and the effects of clouds and
hazes in their atmospheres \citep{moran2018,fauchez2019}.
Observations of the TRAPPIST-1 system with the James Webb Space
Telescope ({\it JWST}) are discussed in detail by
\citet{lustigyaeger2019a}, whose analysis demonstrates that
CO$_2$-rich atmospheres may be detected with $\sim$10 transits, but
aerosol hazes, such as the H$_2$SO$_4$ haze found on Venus, may limit
such detections. The simulations carried out by \citep{fauchez2019}
further discuss the challenges of detecting H$_2$O if the planet is
not in a moist greenhouse state, thus confining the water vapor to the
lower atmosphere.

Several of the TRAPPIST-1 planets lie interior to the HZ in the region
defined as the Venus Zone (VZ), described in detail by
\citet{kane2014e}. GCM models of similar terrestrial planets in high
insolation regimes have indicated a rapid atmospheric evolution toward
a runaway greenhouse scenario, such as the case of Kepler-1649b
\citep{kane2018d}. However, many questions remain regarding the
divergence of the apparent Venus--Earth dichotomy and the relative
effects of insolation flux, water delivery, and the initial conditions
of the interior and atmosphere \citep{kane2019d}. In particular, the
potential diversity of terrestrial planets within the TRAPPIST-1
system provide an opportunity to study possible runaway greenhouse
environments outside of the nominal VZ through {\it JWST} observations
\citep{lincowski2018}. Thus, determining evidence of a post-runaway
greenhouse environment for the TRAPPIST-1 planets would be extremely
insightful for the evolution of terrestrial planets
\citep{lincowski2019,way2020}.

A further consideration is that of atmospheric mass loss of the
TRAPPIST-1 planets, exacerbated by their relatively old age
\citep{burgasser2017,gonzales2019} and high XUV environment
\citep{roettenbacher2017}. For example, the recent discovery of
LHS~3844b \citep{vanderspek2019} was demonstrated through follow-up
observations to have no thick atmosphere \citep{kreidberg2019b},
indicating a volatile-poor formation scenario
\citep{kane2020d}. However, the analysis of transmission spectroscopy
data performed by \citet{moran2018} indicates that the outer (d, e,
and f) planets may have volatile-rich extended
atmospheres. Verification of such extended atmospheres with further
observations is critically important for investigating the interplay
between atmospheric loss due to stellar erosion and on-going
outgassing production of secondary atmospheres \citep{kite2020c}.

In Sections \ref{comb} and \ref{gcm}, we provide predictions of the
phase amplitude due to the reflected light and thermal emission
components. Our modeling does not account for the additional effects
of Doppler beaming and ellipsoidal variation
\citep{loeb2003,zucker2007b}. It is unnecessary to account for these
aspects of the photometric variations since the planets are
terrestrial and will produce negligible beaming and ellipsoidal
amplitudes. Specifically, for the TRAPPIST-1 planets, our calculations
of the Doppler beaming and ellipsoidal variation amplitudes are 1 and
4 orders of magnitude less than the reflected light amplitude,
respectively. For example, in the case of TRAPPIST-1e, the predicted
amplitude of the phase variations due to reflected light is 0.835~ppm
for the atmosphere model (see Table~\ref{tab:flux}). The corresponding
amplitudes of the Doppler beaming and ellipsoidal variations are
0.073~ppm and 0.063~ppb, respectively.

There are several caveats to note with respect to detectability of the
phase signatures described here. The differences in the phase
signatures between the modern Earth-life and Archean-like atmospheres
discussed in Section~\ref{mpcd} could easily become entangled in the
atmospheric signatures, including the effects of cloud distribution
and topography, of other planets within the system, each with their
own phase signatures. In particular, contamination by the residuals
from the inner planets could lead to a similar apparent shift in the
phase maxima that was attributed in Section~\ref{mpcd} to differences
in shortwave and longwave radiation. This problem may be partially
mitigated through a concerted effort to provide a detailed
characterization of the inner planets. The orbital ephemerides of the
system is remarkably well established, and the combination of the
precisely determined planetary orbits with the phase signatures of the
inner planets may allow their effects to be subtracted from
investigations of the phase signatures for planets e and f.

Finally, the detectability of the phase amplitudes presents a
significant observational challenge, even if only a single planet were
present. For example, \citet{wolf2019} demonstrated that measurements
at the level of ppm, and even ppb, may be required for discerning
various aqua-planet scenarios for M dwarf terrestrial
planets. \citet{pidhorodetska2020} provided noise model estimates for
a TRAPPIST-1e spectrum, assuming 85 transits observed with future
exoplanet facilities. The noise model calculations used the detailed
reports for the Habitable Exoplanet Observatory (HabEx) mission
\citep{reporthabex}, the Large UV/Optical/Infrared Surveyor (LUVOIR)
mission \citep{reportluvoir}, and the Origins Space Telescope
\citep{reportorigins}. These noise calculations suggest that HabEx,
LUVOIR, and Origins achieve a 1$\sigma$ noise floor at 5~ppm. This
noise floor is higher than many of the phase amplitudes predicted in
this study for individual planets. Therefore, a more viable goal in
the short-term may be to leverage the precisely determined orbits of
the planets to observe the system during the syzygy events of
planet-planet occultations, described in Section~\ref{comb}. Such a
detection would not easily resolve differences between atmospheres and
topographies due to the degeneracies inherent in a multi-planet fit to
the data, but would indicate the extent of the scattering and
reflective properties of the combined atmospheric profile.

%%%%%%%%%%%%%%%%%%%%%%%%%%%%%%%%%%%%%%%%%%%%%%%%%%%%%%%%%%%%%%%%%%%%

\section{Conclusions}
\label{conclusions}

Exoplanetary science has undergone a significant shift in recent years
toward detailed characterization of terrestrial planets. This has been
enabled by the dramatic rise in discoveries across a broad range of
exoplanet demographics, combined with the prolific development of
ground- and space-based facilities capable of spectroscopy of
planetary atmospheres. Among multi-planet systems, the TRAPPIST-1
system stands out due to its large number of relatively small planets
that reside in a range of insolation environments, allowing
unprecedented studies of comparative planetology. It is therefore
likely that TRAPPIST-1 will be one of the most observed systems with
respect to atmospheric characterization studies.

The process of deep characterization of terrestrial exoplanet
atmospheres requires significant observing time carried out over
multiple wavelength ranges. In addition to the retrieval models
applied to spectroscopic data \citep{barstow2020a}, phase variations
yield additional insights into atmospheric properties. The amplitude
and shape of the variations have a strong wavelength dependence
\citep{sudarsky2005}, also depending on atmospheric composition and
topography \citep{cowan2008,cowan2013b}. Furthermore, the seasonal
variations that may correspond with photometric variability can serve
as a biosignature \citep{olson2018}. Thus, the detection of phase
variations for a system with the astrobiological significance of
TRAPPIST-1 would add complementary information to the overall
characterization of the planets. Even so, the relatively low
signal-to-noise phase signals expected in the face of stellar activity
present significant challenges in the years ahead, motivating
additional effort to distinguish between planetary and stellar
photometric variability \citep{serrano2018}.

In terms of detectable signatures, our simulations show that the
observational prospects of detecting the combined phase amplitude of
several TRAPPIST-1 planets with {\it JWST}, though feasible, will
likely provide limited resolution of the phase change with respect to
the orbital period. A detailed observing campaign that could robustly
use the signatures presented here to distinguish between, for example,
the Archean-like and modern Earth-like scenarios described in
Section~\ref{gcm}, will need to filter out the phase signature from
other planets within the system and also overcome the noise floor
limitations of missions such as HabEx, LUVOIR, and Origins. However,
the prospect of such deep insights into the atmospheric and surface
characteristics of the the TRAPPIST-1 planets motivates further
increasing atmospheric characterization capabilities in the coming
years.

%%%%%%%%%%%%%%%%%%%%%%%%%%%%%%%%%%%%%%%%%%%%%%%%%%%%%%%%%%%%%%%%%%%%

\section*{Acknowledgements}

The authors would like to thank the anonymous referee for the
constructive feedback on the manuscript, and Eric T. Wolf for his
guidance regarding ROCKE-3D and SOCRATES. This research has made use
of the NASA Exoplanet Archive, which is operated by the California
Institute of Technology, under contract with the National Aeronautics
and Space Administration under the Exoplanet Exploration Program. The
results reported herein benefited from collaborations and/or
information exchange within NASA's Nexus for Exoplanet System Science
(NExSS) research coordination network sponsored by NASA's Science
Mission Directorate. This material is based upon work supported by the
National Science Foundation under Grant No. DGE-1644869.

%%%%%%%%%%%%%%%%%%%%%%%%%%%%%%%%%%%%%%%%%%%%%%%%%%%%%%%%%%%%%%%%%%%%

\software{ROCKE-3D \citep{way2017b}, Panoply, astropy}
  
%%%%%%%%%%%%%%%%%%%%%%%%%%%%%%%%%%%%%%%%%%%%%%%%%%%%%%%%%%%%%%%%%%%%

%\bibliographystyle{aasjournal}
%\bibliography{/data/skane/latex/styles/references}

%%%%%%%%%%%%%%%%%%%%%%%%%%%%%%%%%%%%%%%%%%%%%%%%%%%%%%%%%%%%%%%%%%%%

\end{document}